\documentclass[12pt,preprint, epsfig]{aastex}

\newcommand{\myemail}{katagiri@cr.scphys.kyoto-u.ac.jp,enomoto@icrr.u-tokyo.ac.jp}

\shorttitle{Gamma-rays around 1\,TeV from RX J0852.0$-$4622}
\shortauthors{Katagiri, Enomoto et al.}

\begin{document}

\title{Detection of Gamma-rays around 1\,TeV from RX J0852.0$-$4622 by CANGAROO-II}


\author{
H.~Katagiri\altaffilmark{1},
R.~Enomoto\altaffilmark{2},
L.T.~Ksenofontov\altaffilmark{2},
M.~Mori\altaffilmark{2},
Y.~Adachi\altaffilmark{2}, 
A.~Asahara\altaffilmark{1}, 
G.V.~Bicknell\altaffilmark{3}, 
R.W.~Clay\altaffilmark{4}, 
Y.~Doi\altaffilmark{5},   
P.G.~Edwards\altaffilmark{6}, 
S.~Gunji\altaffilmark{5}, 
S.~Hara\altaffilmark{2}, 
T.~Hara\altaffilmark{7}, 
T.~Hattori\altaffilmark{8}, 
Sei.~Hayashi\altaffilmark{9},  
C.~Itoh\altaffilmark{10}, 
S.~Kabuki\altaffilmark{2}, 
F.~Kajino\altaffilmark{9}, 
A.~Kawachi\altaffilmark{2}, 
T.~Kifune\altaffilmark{11},
R.~Kiuchi\altaffilmark{2}, 
H.~Kubo\altaffilmark{1}, 
T.~Kurihara\altaffilmark{8},  
R.~Kurosaka\altaffilmark{2}, 
J.~Kushida\altaffilmark{8} 
Y.~Matsubara\altaffilmark{12}, 
Y.~Miyashita\altaffilmark{8},  
Y.~Mizumoto\altaffilmark{13}
H.~Muraishi\altaffilmark{14}
Y.~Muraki\altaffilmark{12}
T.~Naito\altaffilmark{7},
T.~Nakamori\altaffilmark{1}, 
T.~Nakase\altaffilmark{8}, 
D.~Nishida\altaffilmark{1}, 
K.~Nishijima\altaffilmark{8}, 
M.~Ohishi\altaffilmark{2}, 
K.~Okumura\altaffilmark{2},  
J.R.~Patterson\altaffilmark{4}, 
R.J.~Protheroe\altaffilmark{4}, 
N.~Sakamoto\altaffilmark{5},   
Y.~Sakamoto\altaffilmark{8},   
D.L.~Swaby\altaffilmark{4}, 
T.~Tanimori\altaffilmark{1}, 
H.~Tanimura\altaffilmark{1}, 
G.J.~Thornton\altaffilmark{4}, 
K.~Tsuchiya\altaffilmark{2}, 
S.~Watanabe\altaffilmark{1},  
T.~Yamaoka\altaffilmark{9},  
S.~Yanagita\altaffilmark{15}, 
T.~Yoshida\altaffilmark{15},
T.~Yoshikoshi\altaffilmark{2}}

\altaffiltext{1}
{Department of Physics, Graduate School of Science, Kyoto University, Sakyo-ku, Kyoto 606-8502, Japan; \myemail}
\altaffiltext{2}
{Institute for Cosmic Ray Research, University of Tokyo,  Kashiwa, Chiba 277-8582, Japan}
\altaffiltext{3}
{RSAA, Australian National Univ., ACT 2611, Australia}
\altaffiltext{4}
{Dept.\ of Physics, Univ.\ of Adelaide, SA 5005, Australia}
\altaffiltext{5}
{Department of Physics, Yamagata University, Yamagata, Yamagata 990-8560, Japan}
\altaffiltext{6}
{Institute of Space and Astronautical Science, Japan Aerospace Exploration Agency, Sagamihara, Kanagawa 229-8510, Japan} 
\altaffiltext{7}
{Faculty of Management Information, Yamanashi Gakuin University, Kofu, Yamanashi 400-8575, Japan}
\altaffiltext{8}
{Department of Physics, Tokai University, Hiratsuka, Kanagawa 259-1292, Japan}
\altaffiltext{9}
{Department of Physics, Konan University, Kobe, Hyogo 658-8501, Japan}
\altaffiltext{10}
{Ibaraki Prefectural University of Health Sciences, Ami, Ibaraki 300-0394, Japan} 
\altaffiltext{11}
{Faculty of Engineering, Shinshu University, Nagano, Nagano 480-8553, Japan} 
\altaffiltext{12}
{Solar-Terrestrial Environment Laboratory,  Nagoya University, Nagoya, Aichi 464-8602, Japan} 
\altaffiltext{13}
{National Astronomical Observatory of Japan, Mitaka, Tokyo 181-8588, Japan}
\altaffiltext{14}
{School of Allied Health Sciences, Kitasato University, Sagamihara, Kanagawa 228-8555, Japan}
\altaffiltext{15}
{Faculty of Science, Ibaraki University, Mito, Ibaraki 310-8512, Japan}


\begin{abstract}

We have detected gamma-ray emission at the 6\,$\sigma$ level at energies 
greater than 500\,GeV from the supernova remnant RX J0852.0$-$4622 
(G266.2$-$1.2)
using the CANGAROO-II Imaging Atmospheric Cherenkov Telescope (IACT).
The flux was 0.12 times of that of Crab at 1~TeV.
The signal centroid is consistent with the peak of the X-ray emission
in the north-west rim of the remnant.
\end{abstract}


\keywords{gamma rays: observation --- (stars:) supernovae: individual (RX
J0852.0$-$4622) }


\section{Introduction}
Young supernova remnants (SNRs) have for many years been believed to 
accelerate cosmic rays up to energies around the ``knee''
of the cosmic ray spectrum at $\sim$100\,TeV, 
however direct
observational evidence has only recently been found to support this.
The interactions of such high energy cosmic rays with surrounding material
produce a signature in TeV energy gamma-rays, and so
TeV observations of SNR are therefore
of fundamental importance in understanding the origin of Galactic cosmic rays.
To date, the only SNRs which 
have been reported to 
emit TeV gamma-rays are the shell-type SNRs 
SN1006 \citep{tanimori98}, 
RX~J1713.7$-$3946 \citep{muraishi00, enomoto02b}, and
Cassiopeia~A \citep{aharonian01}.
However, although RX J1713.7$-$3946 has been
detected by two groups \citep{hofmann04},
there have as yet been no confirming reports for Cassiopeia~A,
and for SN1006, a recent upper limit is significantly
below the originally reported flux \citep{hofmann04}.
Therefore, in addition to further observations of these sources, 
it is necessary to search for further examples of such SNRs.

RX J0852.0$-$4622 (G266.2$-$1.2) is a SNR located 
(in projection) 
at the southeast corner of the Vela SNR.
It was discovered by \cite{aschenbach98} using data from the
 {\it ROSAT} All-Sky Survey.
It is one of the few known SNRs 
which display strong non-thermal X-ray emission \citep{tsunemi00, slane01}.
In addition, the 1157\,keV line of $^{44}$Ti was detected by COMPTEL \citep{iyudin98}.
From a consideration of the lifetime of $^{44}$Ti ($\sim$90\,yr), 
the X-ray flux, and the angular size of the X-ray emission, $\sim$2$^\circ$,
the age and distance from the Earth were 
estimated to be 500--1100\,yr and 80--500\,pc, 
respectively \citep{aschenbach99}.
An independent estimate of the distance, of 1--2\,kpc,
was provided by \cite{slane01}
from a simple scaling of the column density,
which suggested that the SNR was at least 
several times more distant than the Vela SNR.  
{\it ASCA} hard X-ray images revealed a shell-like morphology
with a featureless spectrum well described by a power-law 
\citep{tsunemi00, slane01}.
Radio emission also has a non-thermal spectrum \citep{combi99, duncan00}.
Both the X-ray and radio emission are brightest in the north-west rim of the 
SNR.
These data suggest that particles may be being accelerated to $\sim$100\,TeV.
At these energies electrons can give rise to TeV gamma-rays
through inverse Compton scattering, and protons can interact with 
surrounding material to produce TeV gamma-rays via $\pi^0$ decay.

Furthermore, RX J0852.0$-$4622 is, like RX J1713.7$-$3946,
an extended SNR at X-ray energies.
The relatively large angular size enables us to study the 
detailed morphology 
even with the existing angular resolution.

We have observed RX J0852.0$-$4622 with the
10\,m diameter
CANGAROO-II Imaging Atmospheric Cherenkov Telescope (IACT).
The telescope \citep{kawachi01}
detects gamma-rays above several hundred GeV
by reconstructing optical
Cherenkov images generated by relativistic secondary particles in
the cascades produced when high-energy gamma-rays (and background cosmic rays)
interact with the Earth's upper atmosphere. 
Parametrizing the optical images enables the 
direction and energy of gamma-ray events to be inferred.
The telescope is located near Woomera, South Australia 
(136$^\circ$47$'$\,E, 31$^\circ$06$'$\,S), from which
RX J0852.0$-$4622
culminates at $\sim$15$^\circ$ from the zenith, enabling 
observations with a lower energy threshold.
The angular resolution of CANGAROO-II was estimated to be 0.30--0.39$^\circ$
(2.6--3.4\,pc at a distance of 0.5\,kpc) with an energy
threshold of 500\,GeV depending on the observing conditions
and the spectrum.

\section{Observations}

The observations were carried out between 2001 December~12 
and 2002 February~15 (19 nights), and 
between 2003 January~5 and February~28 (25 nights). 
The telescope tracked the peak of the X-ray emission, in
the north-west rim of the SNR
($\alpha =132.25^\circ$, $\delta =-45.65^\circ$, J2000 coordinates).
The field of view (FOV) of the camera is 2.76$^\circ \times$2.76$^\circ$.
The brightest star in the FOV, SAO 220422, has a visual magnitude of 4.1,
and this region is brighter than our typical on-source FOV. 
As a result, we used a higher trigger threshold,
requiring 5 triggered pixels rather than the usual 3 \citep{cito03}.
The recently installed lights several kilometers from
the telescope were expected to have 
little effect on these observations as the telescope
was pointed close to the zenith \citep{cito03}, and
this was confirmed from the azimuthal angle dependence of the shower rate
(see \S\,3).
Each night was divided into two or three periods, i.e., ON--OFF,
OFF--ON--OFF, or OFF--ON observations. ON-source observations were timed
to contain the meridian passage of the target, 
as was done by \cite{enomoto02b}.
In total, 5900 min.\ of ON- and 5300 min.\ of OFF-source data
were obtained.

\section{Analysis}

First, `cleaning' cuts on camera images were applied,
requiring (0.115$^\circ$-square) pixel pulse-heights of greater than
3.3 photoelectrons, and Cherenkov photon arrival times within $\pm$50\,ns
of the median arrival time.
Clusters of at least five adjacent triggered pixels 
(rather than the usual four-pixel cut)
were required in each event
to minimize the effects of the bright star field.
After these pre-selection cuts, which reduced
events due to background light were reduced by 99\%,
the shower rate was stable on a
run-to-run basis for observations in the same year.
The systematic difference of the run-by-run acceptance within the
same year is 
expected to be less than 12\%.
The ON/OFF shower rate differences in 2002 and 2003 were $-8\pm15$\% and 
$-1\pm13$\%, respectively.
By examining the event rates within each run 
we were able to reject periods affected by cloud, 
dew forming on the mirrors,
instrumental abnormalities,
etc.
Only data taken at elevation angles greater than 60$^{\circ}$ were accepted.
After these cuts, 4300 min.\ of ON- and 3900 min.\ of OFF-source
data survived.

Trigger rates for each pixel per 700\,$\mu$s were monitored by a 
scaler circuit in real-time and recorded each second.
These data were used to exclude `hot' pixels (generally due to 
the passage of a star through the FOV of a pixel) in off-line
analysis.  
Hillas parameters were then calculated to discriminate gamma-rays
from cosmic rays based on the image shape and orientation  \citep{hillas85}.
Further, in each year's data, we masked a small number of pixels which 
 showed deformed ADC spectra, possibly due to a hardware fault.
Discrimination of the cosmic ray background from gamma-rays
was carried out using the likelihood method of \cite{enomoto02a}.

\section{Results}

\begin{figure}
\plotone{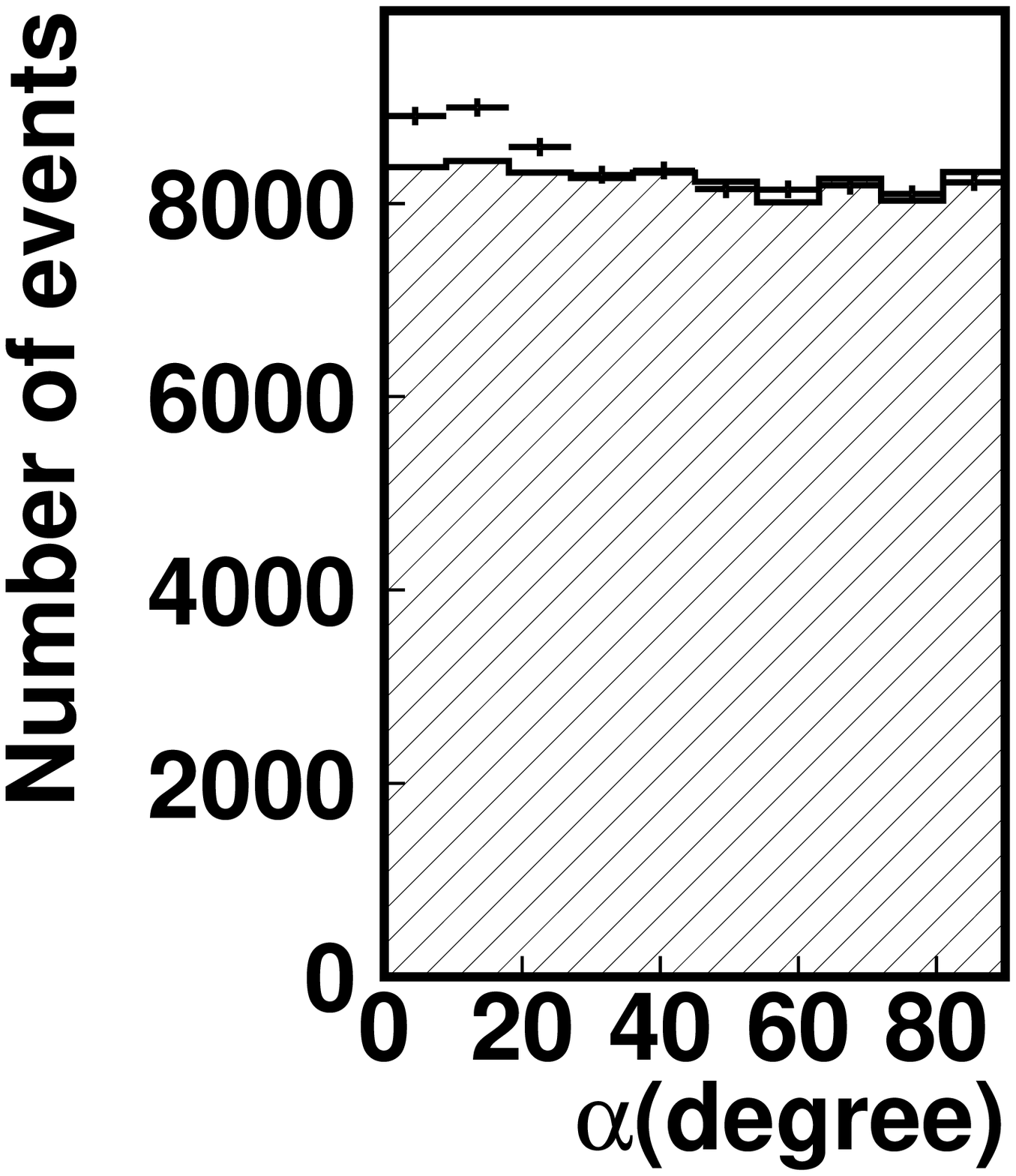}
\caption{Distributions of $\alpha$ (image orientation angle)
for the combined data in 2002 and 2003.
The points with error bars show the ON-source data 
and the hatched histogram is the OFF-source data, which was normalized
to the ON-source data using the number of the events with $\alpha>27^\circ$.   }  
\label{fig1}
\end{figure}

The resulting distributions of the image orientation angle, $\alpha$,
for the combined data in 2002 and 2003 
are shown in Fig.~\ref{fig1}.
The normalizations between the ON- and
OFF-distributions were carried out using data with
$\alpha>27^\circ$.
The numbers of excess events ($\alpha <18^\circ$)
were  $530\pm120$ (in an observation time of 2100 min.), $540\pm140$
(2200 min.), and $1080\pm180$ (4300 min.) in 2002, 2003, 
and the combined data, respectively.
The excess rates for 2002 and 2003 were similar to each other.
Nightly signal rates were also checked during both years.
The largest deviations occurred 
with a rate $2.5\pm 0.9$ times larger than the average for 2002 
and  $3.0\pm 1.0$ times for 2003, respectively.
These are not unexpected statistically, and therefore 
there is no evidence of time-variability in the TeV emission.

\begin{figure}
\plotone{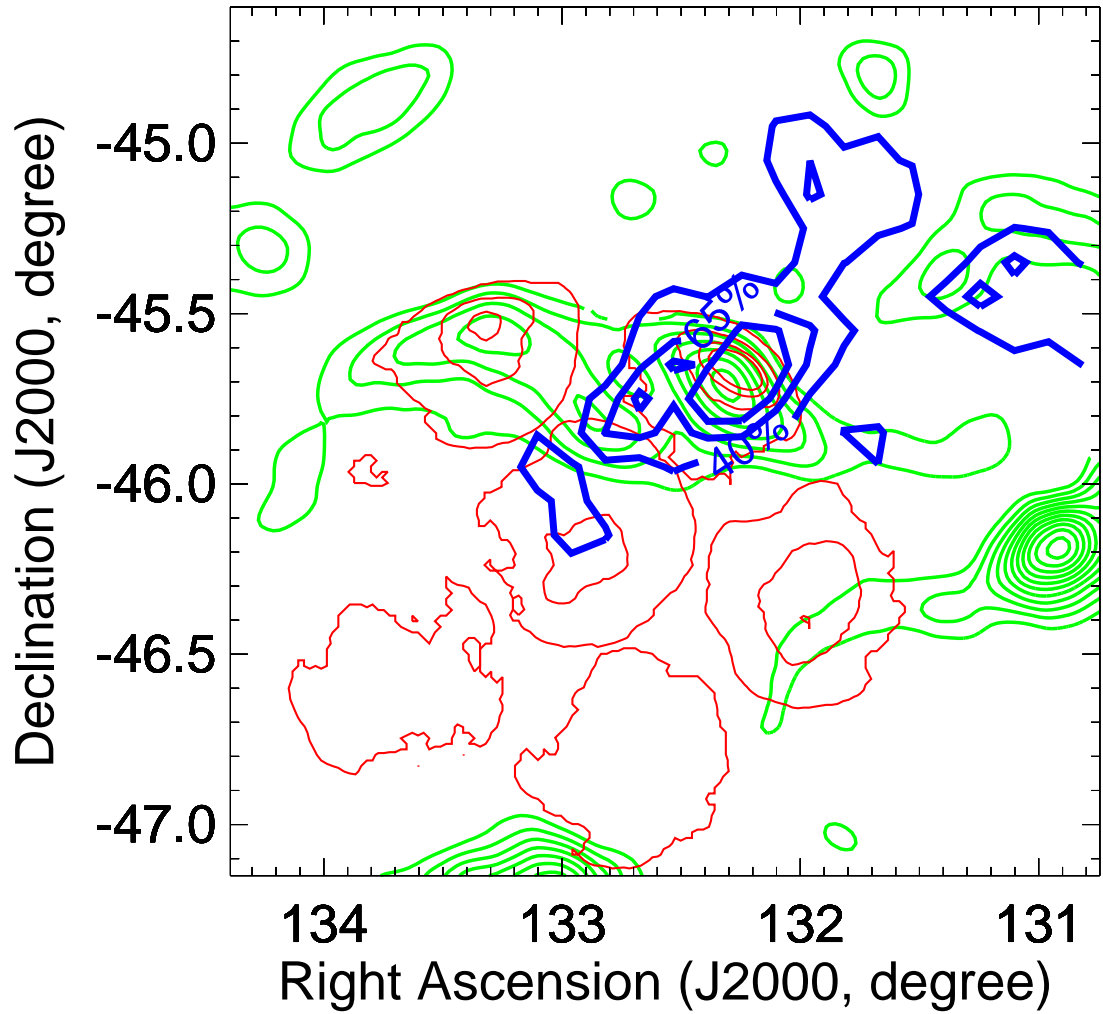}
\caption{ 
The ``significance map'' obtained by the CANGAROO-II telescope 
is shown by the blue contours. 
The telescope tracking center is at the north-west rim of RX J0852.0$-$4622. 
The red contours with levels at 20\%, 45\%, 65\%, and 80\% 
are an {\it ASCA} GIS image.
The green contours show the 4850\,MHz radio emission.}
\label{fig2}
\end{figure}

To check on the spatial distribution of the signal
we derived the ``significance map'', shown by the blue contours in 
Fig.~\ref{fig2}.
The contours were
calculated from the distribution of the detection significance
determined at each location from the difference in the $\alpha$ plots
(ON- minus OFF-source histogram) divided by the statistical errors.
The centroid is consistent with position
of the X-ray maximum, within
our possible systematic uncertainty of 0.1$^\circ$.
The 
region enclosed by the 65\% (of the peak TeV significance) contour 
is elliptical with a 
semi-minor axis of 0.19$^\circ$ (NE-SW) 
and a semi-major axis of 0.35$^\circ$ (NW-SE), while
our angular resolution was estimated to be 0.30--0.39$^\circ$.
The acceptance of the CANGAROO-II telescope 
decreases smoothly with offset from the tracking center, falling to
50\% at a 0.9$^\circ$ offset, i.e. at the edge of the trigger region.
It is difficult to obtain the reliable upper limits on the two neighboring 
X-ray enhancements in the rim
since they are near the edge of the trigger region.

After correcting for this acceptance, the differential fluxes 
listed in Table~\ref{table1} were derived.
As both statistical and systematic errors are included,
the energy bins overlap somewhat, particularly at low energies. 
The systematic uncertainty for the energy determination
($\sim$20\%) dominates the errors in the energies.
The Spectral Energy Distributions (SEDs),
derived from radio, X-ray, MeV gamma-ray, and our TeV gamma-ray observations,
are plotted in Fig.~\ref{fig3}.
The hatched area indicates the allowed region, where the $\chi^2$ for the 
power-law fitting $<$ $\chi^2_{\rm min}+1$ including energy uncertainty, 
and $\chi^2_{\rm min}$ is the $\chi^2$  for the best fit. 
The values are listed in Table \ref{table1}. 
The derived spectrum has a power-law index
of $-4.3^{+1.7}_{-4.4}$, the errors of which come 
predominantly from the uncertainty in energy.

The derived spectrum from RX J0852.0$-$4622 seems rather steep.
The photon index of $-$2.6, however, is within 1\,$\sigma$ level due to
 the large uncertainty on the energy determination.
This indicates that we cannot firmly conclude
that the energy spectrum as soft as spectral index of $-$4.3.
Recently CANGAROO-II collaboration has reported a similar steep spectrum
from the Galactic Center \citep{tsuchiya04}.
This spectrum is different from that reported by H.E.S.S. collaboration
\citep{aharonian04a}.
Applying the same procedure to the Galactic Center result,
the allowed spectral index range became $-$4.6$^{+1.2}_{-5.0}$.
No further answer on this is found in this stage.
Time variability might be one of solution
\citep{neronov04}.
On the other hand,
the difference in the energy spectrum from the SNR RX J1713.7$-$3946
between the CANGAROO-II and the H.E.S.S.
is 2.2\,$\sigma$ level \citep{aharonian04b}.
These complex situations encourage
the further observations and analysis of the stereoscopic data using IACTs.

Our Monte Carlo simulations predict, even for a point-source,
that the $\alpha$ distribution of gamma-ray events for this soft spectrum 
should have a broader peak than that for a Crab-like spectrum.
Therefore we cannot conclude from the experimental data whether 
this source is extended or point-like.

Various checks on the signal yield and position were carried out, by
varying thresholds, clustering cuts, Hillas parameter values, etc:
these yielded consistent fluxes within the systematic
errors given in Table \ref{table1}. 
Also, Crab nebula data were analyzed with the same code, with
the derived flux and morphology consistent with 
previous measurements and the point-spread function, respectively.
The distributions of the Hillas parameters for the excess events
were checked and found to be to be consistent with Monte Carlo
simulations for gamma-rays. The OFF-source data
were compared with the Monte Carlo simulations for protons,
and found to be consistent.

\section{Discussion}
\begin{figure}
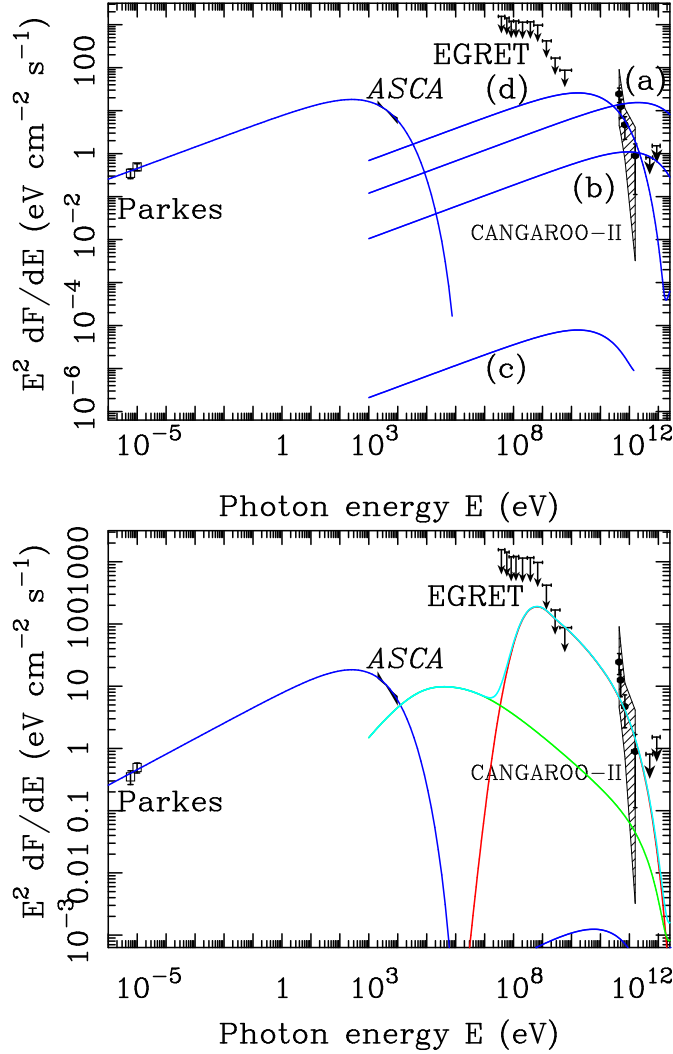

\epsscale{.54}
\includegraphics[scale=0.40,angle=-90]{f3a.eps}
\includegraphics[scale=0.40,angle=-90]{f3b.eps}
\centering

\caption{ 
Spectral energy distribution of the north-west rim of RX J0852.0$-$4622.
The points with error bars and the arrows in the TeV region are the 
fluxes and the 2$\sigma$ upper limits, respectively, from our observations.
The hatched area indicates the allowed region, 
where the $\chi^2$  for the power-law fitting 
$<$ $\chi^2_{\rm min}+1$ including
energy uncertainty, and $\chi^2_{\rm min}$ is the $\chi^2$  for the best fit.
The arrows around 100\,MeV show the 2\,$\sigma$ upper limits obtained using 
the EGRET data of the diffuse emission.
They were derived from the count rate plus twice the standard deviation based on the archival data of the EGRET gamma-ray intensity map, which was integrated over the area considering the angular resolutions of EGRET and CANGAROO.
The open squares with the error bars show
the Parkes radio data \citep{duncan00}. 
The black filled areas show the {\it ASCA} X-ray data \citep{slane01}.
The red line shows the estimation for $\pi^0$ gamma-rays, the details
 of which are given in the text. 
The blue lines shows the spectra of the
various synchrotron/inverse Compton (IC) models.
The green line shows the bremsstrahlung emission from electrons.
The light blue line shows the sum of the $\pi^0$, the inverse Compton, and the bremsstrahlung emissions.
\label{fig3}}
\end{figure}

The gamma-rays around 1\,TeV are most likely produced by one of
two mechanisms. 
We consider first the synchrotron/inverse Compton (IC) model.
The spectra obtained using this model are shown in Fig.~\ref{fig3} (upper). 
This model has the following parameters: the magnetic field ($B$), the spectral
index of electrons ($\gamma$), and the maximum 
electron energy ($E_{\rm e,max}$).
The cutoff energy of the synchrotron emission is proportional to
$E_{\rm e,max}^2 B$,
which was constrained by the X-ray and the radio data.
The cutoff of the IC emission 
is proportional to $E_{\rm e,max}^2$. 
The ratio of the peak of the synchrotron component 
and the peak of the IC component in the SED 
is proportional to $B^2$.
If $B$ is too small, $E_{\rm e,max}^2$ is too high to 
explain our data, as shown by (a) in Fig~\ref{fig3}.
The lower limit of $B$ in this model is $\sim$12\,$\mu$G  
as shown by (b) in Fig.~\ref{fig3}. This model cannot explain our data.

A more complex model in which the synchrotron and IC emissions come from
different zones (a two-zone model) was considered by \cite{aharonian97}. 
At first, we assume the same electron spectra and $B$ 
with different zone
sizes. 
The lines(c) and (d) in Fig.~3 show the IC emissions with size ratios 
(V$_{\rm TeV}$/V$_{\rm X \textrm{-} ray}$) of 1 and $\sim 10^5$, respectively. 
The flux ratio between lines (c) and (d) 
corresponds roughly to the size ratio.
This model requires a high size ratio and a strong magnetic field 
($\sim 1.6$~mG).
{\it Chandra} observations revealed
small-scale structures in the rim of SN1006 \citep{bamba03}:
similar high resolution observations (in both X-rays 
and TeV gamma-rays) of RX J0852.0$-$4622 
will help constrain the size ratio and refine this two-zone model.
Stereo observations in the TeV region are required.

The other possible origin for the gamma-rays is $\pi^0$ decay.
The estimated flux from $\pi^0$ decays is shown
 by the red line in Fig.~\ref{fig3} (lower).
The emissions from the electrons were also considered 
using the electron/proton ratio ($K_{\rm ep}$).
Bremsstrahlung emission, shown by the green line,
is constrained by the {\it ASCA} data.
Therefore $K_{\rm ep}$ was limited to be at most 4$\times 10^{-4}$.
As a result, both the flux of the inverse Compton emission and that of 
the bremsstrahlung were too small at TeV energies to explain our data.
The sum of the $\pi^0$ emission, the inverse Compton emission and the bremsstrahlung is shown by the light blue line.
The spectrum of accelerated protons is assumed to be a
  power-law with an exponential cutoff.
The $\pi^0$ model has the following parameters: the spectral index of
the protons, the maximum energy of protons ($E_{\rm p,max}$), and
$A\equiv(E_0/10^{50}{\rm erg})(n_0/{\rm protons}\,{\rm cm}^{-3})(d/0.5{\rm kpc})^{-2}$, 
respectively.
Here,
$E_0$, $n_0$, and $d$ are the total energy of
protons in the region from which gamma-rays were detected, the number
density of protons where the interactions occur, and the distance from
the earth, respectively.
The spectral index and the maximum energy of protons were assumed to be 
those of electrons, i.e., $-$2.5 and 8\,TeV, respectively.
For the best-fit spectrum $A=$4.9 was obtained.
Assuming the distribution of the energy of the accelerated cosmic rays is 
isotropic, 
the total energy of accelerated cosmic rays in the whole SNR 
is proportional to the ratio V$_{\rm tot}$/V$_{\rm TeV}$, 
where V$_{\rm tot}$ is the total volume. 
This ratio was estimated to be 46 from the ``significance map'' 
with the assumption of spherical symmetry.
The range of $E_0$(V$_{\rm tot}$/V$_{\rm TeV}$) of 
$10^{48}$--$10^{50}$ergs corresponds to $n_0$ of
23000--230, which are typical values for molecular clouds.
Thus, the model of $\pi^0$ decays can naturally explain the multi-band
spectrum.

The molecular distribution toward the whole extent of the Vela SNR
has been mapped by \cite{moriguchi01}.
As RX J0852.0$-$4622 is embedded (in projection) in the Vela SNR,
it is not possible to uniquely associate the detected molecular
clouds with one or other SNR.
The general anti-correlation
of the X-rays and $^{12}$CO with the Vela SNR suggests a strong
interaction between this SNR and the interstellar medium,
but does not rule out the existence of 
molecular clouds around RX J0852.0$-$4622, 
encouraging the further analysis.

\acknowledgments
This work was supported by a Grant-in-Aid for Scientific Research by
the Japan Ministry of Education, Culture, Sports, Science and Technology 
(MEXT) of Japan,  
 the 21st Century COE ``Center for Diversity and Universality in Physics'' from 
MEXT,
 the Australian Research Council, ARC Linkage Infrastructure Grant 
LE0238884, Discovery Project Grant DP0345983, and JSPS Research Fellowships.
We thank the Defense Support Center Woomera and BAE Systems.
We are indebted to Dr. H.~Matsumoto for obtaining the X-ray spectrum.


\clearpage


\clearpage
\clearpage

\clearpage

\begin{table}   
\begin{center}
\caption{Differential fluxes.}
\label{table1}
\begin{tabular}{cccc} 
\hline          
\hline               
Mean energy  &  Flux  & Lower bound$^*$ & Upper bound$^*$ \\
of bin [TeV]  & [ph/cm$^2$/s/TeV] &  &  \\
\hline
 0.48$\pm$0.11 & (9.43$\pm$3.49)$\times$$10^{-11}$ & 1.99$\times$$10^{-11}$ & 3.01$\times$$10^{-10}$ \\
 0.55$\pm$0.12 & (3.85$\pm$1.78)$\times$$10^{-11}$ & 1.03$\times$$10^{-11}$ & 1.28$\times$$10^{-10}$ \\ 
 0.79$\pm$0.18 & (8.54$\pm$4.63)$\times$$10^{-12}$ & 1.47$\times$$10^{-12}$ & 1.84$\times$$10^{-11}$ \\
 1.93$\pm$0.48 & (2.80$\pm$2.45)$\times$$10^{-13}$ & 1.35$\times$$10^{-15}$ & 1.22$\times$$10^{-12}$ \\
 5.84$\pm$1.51 &  3.18$\times$$10^{-14 ~ **}$ & ---  & ---  \\
 10.38$\pm$3.15 &  1.91$\times$$10^{-14 ~ **}$  & --- & ---   \\
\hline
\hline
\end{tabular}
\end{center}
\tablenotetext{*}{These were estimated
 where the $\chi^2$  for the power-law fitting $<$ $\chi^2_{\rm min}+1$ including
energy uncertainty, where $\chi^2_{\rm min}$ was the $\chi^2$  for the best fit.}
\tablenotetext{**}{These are 2\,$\sigma$ upper limits.}
\end{table}

\end{document}